\documentclass[aip,jcp,reprint,twocolumn,nofootinbib,showkeys]{revtex4}

\pdfoutput=1

\usepackage[utf8x]{inputenc}
\usepackage{titlesec}

\usepackage{microtype,lipsum}
\usepackage{enumitem}
\usepackage{amsmath,amssymb,bm,mathtools,dsfont,mathtools}
\usepackage{graphicx}
\usepackage[toc,page]{appendix}

\usepackage{ulem}

\renewcommand{\vr}{\bm{r}}
\newcommand{\vu}{\bm{u}}
\newcommand{\vv}{\bm{v}}
\newcommand{\vom}{\bm{\omega}}

\newcommand{\vR}{\bm{R}}
\newcommand{\vF}{\bm{F}}
\newcommand{\vT}{\bm{T}}

\newcommand{\Pe}{\textrm{Pe}}
\newcommand{\mixr}{\chi_A}

\renewcommand{\d}{\textrm{d}}

\newcommand{\twocomparr}[2]{\left( \begin{array}{c} #1 \\ #2 \end{array}\right)}

\usepackage{ulem}
\usepackage{color}
\newcommand{\GP}[1]{\textcolor{black}{#1}}

\titleformat{\section}[wrap]
{\normalfont\itshape} 
{\thesection.}{0.5em}{}

\titlespacing{\section}{12pc}{1.5ex plus .1ex minus .2ex}{1pc}

\begin{document}


\title{
{\normalfont \textit{\footnotesize This is a preprint of an article published by Taylor \& Francis in Molecular Physics on July 11, 2018, \\[-.2cm] available online: http://www.tandfonline.com/10.1080/00268976.2018.1496291.}}\\
\vspace{0.2cm}
{\small Invited Contribution to the Special Issue of Molecular Physics in Honor of Daan Frenkel} \\
\vspace{0.2cm}
Binary pusher-puller mixtures of active microswimmers and their collective behavior
}

\author{Giorgio Pessot}
\email{giorgio.pessot@hhu.de}
\affiliation{Institut f\"ur Theoretische Physik II: Weiche Materie,
 Heinrich-Heine-Universit\"at D\"usseldorf, D-40225 D\"usseldorf, Germany}

\author{Hartmut L\"owen}
\email{hartmut.loewen@uni-duesseldorf.de}
\affiliation{Institut f\"ur Theoretische Physik II: Weiche Materie,
 Heinrich-Heine-Universit\"at D\"usseldorf, D-40225 D\"usseldorf, Germany}

\author{Andreas M.\ Menzel}
\email{menzel@hhu.de}
\affiliation{Institut f\"ur Theoretische Physik II: Weiche Materie,
 Heinrich-Heine-Universit\"at D\"usseldorf, D-40225 D\"usseldorf, Germany}

\date{\today}

\begin{abstract}
Microswimmers are active particles of microscopic size that self-propel by setting the surrounding fluid into motion.
According to the kind of far-field fluid flow that they induce, they are classified into pushers and pullers.
Many studies have explored similarities and differences between suspensions of either pushers or pullers, but the behavior of mixtures of the two is still to be investigated.
Here, we rely on a minimal discrete microswimmer model, particle-resolved, including hydrodynamic interactions, to examine the orientational ordering in such \GP{binary} pusher-puller mixtures.
In agreement with existing literature, we find that our monodisperse suspensions of pushers do not {show alignment}, whereas \GP{those of solely} pullers spontaneously develop ordered collective motion.
By continuously varying the composition of the binary mixtures, starting from pure puller systems, we find that ordered collective motion is largely maintained up to pusher-puller composition ratios of about 1:2.
Surprisingly, pushers when surrounded by a majority of pullers are more tightly aligned than indicated by the average overall orientational order in the system.
Our study outlines how orientational order can be tuned in active microswimmer suspensions to a requested degree by doping with other species.
\end{abstract}



\maketitle

\section{Introduction}\label{introduction_section}
The field of self-propelled particles and active matter has developed into a prime area to study the properties of non-equilibrium systems. Examples that have been addressed in detail are the statistics of the migrational behavior of individual self-propelled agents involving stochastic fluctuations \cite{howse2007self,ten2011brownian,romanczuk2012active, cates2012diffusive,bechinger2016active} or of their collective motion, including their dynamical phase behavior \cite{vicsek1995novel,chate2008collective, fily2012athermal,stenhammar2013continuum, mognetti2013living,romensky2014tricritical, speck2014effective,menzel2015tuned, menzel2016way,bechinger2016active,siebert2017phase}. 

The vast majority of studies on the collective behavior in this field concentrates on monodisperse systems. To some extent, mixtures of active and passive particles have been investigated.
This concerns the collective behavior of mixtures of self-propelled and passive rods, in which, for instance, laning of the active rods in the passive background is observed \cite{mccandlish2012spontaneous}.
Moreover, the separation into dense and more dilute regions in mixtures of active and passive spherical particles was addressed \cite{takatori2015theory, kummel2015formation, stenhammar2015activity, wysocki2016propagating}, as was the coarsening of crystal domains when systems of passive particles were doped by active agents \cite{van2016fabricating}.
A related topic is the study of mixtures of particles of different temperatures \cite{grosberg2015nonequilibrium, weber2016binary, smrek2017small}.

Investigations on mixtures of different types of active particles are exceptions.
For instance, multi-species swarms of microorganisms were addressed \cite{ben2016multispecies}, predator-prey scenarios were analyzed \cite{mecholsky2010obstacle,sengupta2011chemotactic}, mixtures of active rotors of opposite sense were considered \cite{nguyen2014emergent} including doping by passive particles \cite{yeo2016dynamics}, a stochastic description of mixtures of particles of different activity was outlined \cite{wittmann2018effective}, the alteration of the transition to polarly ordered collective motion with increasing polydispersity of the aligning self-propelled agents was investigated \cite{guisandez2017heterogeneity}, and the mutual support between different species in their orientational ordering and collective motion was studied in the context of imposed alignment interactions \cite{menzel2012collective}.
Mostly, these works concentrate on ``dry'' systems of self-propelled particles, not taking into account the role of a surrounding medium between the individual agents. 

Active microswimmers represent one special type of such self-propelled particles \cite{elgeti2015physics, menzel2015tuned}.
These objects are suspended in a surrounding fluid.
Examples are given by artificial colloidal Janus particles that propel by localized asymmetric concentration or temperature gradients induced in their environment \cite{howse2007self, jiang2010active, buttinoni2012active}.
Biological microswimmers are found in nature in the form of mechanically propelled bacteria or algae \cite{polin2009chlamydomonas,min2009high}.
Their mechanism of self-propulsion sets the surrounding fluid into motion.
As a first coarse classification, one may distinguish between two different types of active microswimmers.
If, to leading order, the induced flow field describes fluid pushed out along the propulsion axis and is dragged in from the sides, the swimmer is called a \textit{pusher} \cite{lauga2009hydrodynamics}.
In the opposite case of fluid being pulled inwards towards the swimmer along the propulsion axis and ejected to the sides, it is classified as a \textit{puller} \cite{lauga2009hydrodynamics}.
Via these induced fluid flows, hydrodynamic interactions \cite{batchelor1972hydrodynamic, batchelor1976brownian, dhont1996introduction, reichert2004hydrodynamic, lauga2009hydrodynamics, malgaretti2012running, bechinger2016active} arise between the individual swimmers that can affect the overall collective behavior \cite{nash2010run, ezhilan2013instabilities, zottl2014hydrodynamics, menzel2014active, matas2014hydrodynamic,hennes2014self, blaschke2016phase, menzel2016dynamical, hoell2017dynamical}.
Due to the small dimensions of microswimmers, the relevant fluid flows are typically characterized by low Reynolds numbers \cite{purcell1977life}. 

In the present work, we combine the two aspects described above. That is, we study mixtures of simplified active model microswimmers that hydrodynamically interact with each other through self-induced fluid flows in suspension. 
More precisely, we investigate binary mixtures of pusher- and puller-type swimmers. 
We concentrate on the microscopic swimmer-scale level, explicitly taking into account the hydrodynamic interactions on this scale. The swimmers are resolved individually in a discretized description using a minimal swimmer model \cite{menzel2016dynamical, hoell2017dynamical}. On this basis, we evaluate the global orientational behavior.

The arising orientational ordering in crowds of microswimmers due to hydrodynamic interactions has been analyzed before for suspensions of either pushers or pullers separately \cite{evans2011orientational, saintillan2011emergence, ezhilan2013instabilities, alarcon2013spontaneous}. 
Here we study this effect in mixtures of the two types. 
In our computer simulations \cite{frenkel2001understanding} we find, for instance, that pushers surrounded by a majority of pullers exhibit tighter orientational ordering than the surrounding pullers.
Underlying details like possible intermittent or spatially localized orientational ordering of the swimmers can be analyzed accordingly in more detail in the future.

Below, we proceed in the following way. First, we describe the equations of motion for our suspended pusher and puller microswimmers.
Afterwards, we analyze the collective behavior of binary mixtures of the two swimmer species for varying amounts of mixing ratio.
In this context, also the impact of temperature and area fraction is addressed.
Some conclusions are added in the end.



\section{Model}
We consider a total of $N$ self-propelled microswimmers, $N_A$ of which are pushers and $N_B= N -N_A$ are pullers, with positions $\vR_i = (R_i^x, R_i^y)$ and normalized orientational vector $\vu_i =(u_i^x, u_i^y)$ ($i=1\dots N$).
For an undisturbed swimmer, $\vu_i$ coincides with its propulsion direction.
All positions and orientations are confined to a two-dimensional plane generated by the directions $\widehat{x}$ and $\widehat{y}$.
Still, three-dimensional hydrodynamic interactions apply.
For brevity, we introduce the multi-dimensional vectors $\vR$ and $\vu$, the components of which are given by the positional and orientational coordinates, respectively, of all swimmers.
Moreover, in the two-dimensional plane, the orientational vector $\vu_i$ of each swimmer can be represented by its angle $\theta_i$ with the $x$-axis such that $\vu_i = (\cos \theta_i , \sin \theta_i )$.
In a similar fashion, we denote by $\bm{\theta}$, $\vv$, and $\vom$ the multi-dimensional vectors containing the angles, the linear, and the angular velocities of all swimmers.
The microswimmers are confined to a two-dimensional periodic square box of area $A$.

In the low-Reynolds-number regime of active microswimmers, dissipation dominates, and the motion is governed by an overdamped, stochastic Langevin equation.
It is to be integrated forward in time according to Stratonovich calculus \cite{vankampen1992stochastic}.
Here we employ a simple Euler integration scheme at the cost of introducing a ``spurious drift'' term \cite{ermak1978brownian, hennes2014self}.

By integrating the Langevin equation over a small time interval $\d t$, 
we obtain the following expressions for the discrete increments $\d \vR$ and $\d \bm{\theta}$ \cite{makino2004brownian, decorato2015hydrodynamics}
\begin{align}
\twocomparr{\d \vR}{\d \bm{\theta}} &=  \twocomparr{\vv_{det}}{\vom_{det}} \d t  + \mathds{H} \cdot\bm{\xi} \sqrt{\d t} \label{langevin_eq}
\end{align}
with the deterministic linear and angular velocities
\begin{align}
 \twocomparr{\vv_{det}}{\vom_{det}} = & \ \mathds{M} \cdot \twocomparr{\vF}{\vT} + \mathds{A} \cdot \twocomparr{\vu}{\bm{0}} + \twocomparr{\partial_{\vR}}{\partial_{\bm{\theta}}} \cdot \mathds{D} \label{v_determ}
\end{align}
as well as the mobility and active mobility matrices
\begin{align}
  \mathds{M} = \twocomparr{\mathds{M}^{tt} \ \mathds{M}^{rt}}{\mathds{M}^{tr} \ \mathds{M}^{rr}}
  \ \mbox{and} \
  \mathds{A} = \twocomparr{\mathds{A}^{tt} \ \mathds{A}^{rt}}{\mathds{A}^{tr} \ \mathds{A}^{rr}}.
\end{align}

The first term on the right-hand side of Eq.~(\ref{v_determ}) determines the contributions of the conservative forces $\bm{F}$ and torques $\bm{T}$ to the deterministic velocities $\vv_{det}$ and angular velocities $\vom_{det}$.
The second term includes the contribution of self-propulsion along each particle axis $\vu_i$.
The last term is the spurious drift \cite{ermak1978brownian}, i.e., the divergence of the diffusion matrix $\mathds{D} = k_BT \mathds{M}$.
In the case of our hydrodynamic interactions, see below, the drift term vanishes \cite{ermak1978brownian}.
Finally, the matrix $\mathds{H}$ is obtained by Cholesky decomposition \cite{press2007numerical} to satisfy $\mathds{H} \cdot \mathds{H}^{T}=2\mathds{D}$.
The components of the vector $\bm{\xi}$ are uncorrelated Gaussian random numbers of zero mean and of variance unity.

Thus, we obtain the correct deterministic mean displacements
\begin{align}
\left\langle \left( \begin{array}{c}  \d \vR\\ \d \bm{\theta} \end{array} \right) \right\rangle &= \mathds{M}\cdot \left( \begin{array}{c}  \bm{F} \\ \bm{T} \end{array}\right) \d t
+ \mathds{A} \cdot \twocomparr{\vu}{\bm{0}} \d t,
\end{align}
and, in the absence of deterministic driving forces and torques, the correct mean squared displacements
\begin{align}
\langle \d \vR\ \d \vR \rangle &= 2 k_BT \mathds{M}^{tt} \d t, \nonumber \\
\langle \d \vR\ \d \bm{\theta} \rangle &= 2 k_BT \mathds{M}^{tr} \d t, \nonumber \\
\langle \d \bm{\theta}\ \d \bm{\theta} \rangle &= 2 k_BT \mathds{M}^{rr} \d t \label{mdispl}
\end{align}
($i,j=1\dots N$) that reproduce the correct time evolution of the corresponding Smoluchowski equation \cite{vankampen1992stochastic, ermak1978brownian}.



\section{Details of the hydrodynamic and steric swimmer interactions}\label{sect_activity}
Hydrodynamic couplings between the swimmers are considered on the Rotne-Prager level \cite{menzel2016dynamical, dhont1996introduction, hoell2017dynamical}.
Each swimmer consists of a spherical body of no-slip surface conditions for the surrounding fluid.
Non-hydrodynamic forces and torques acting on such a swimmer body are transmitted to the surrounding fluid, set it into motion, and in this way affect the motion of all other swimmer bodies, see Eq.~(\ref{v_determ}).
Examples are conservative forces originating from steric repulsion or forces and torques resulting from external fields.
For the $i$-th swimmer, the corresponding components of $\mathds{M}$ are
\begin{align}
&\textrm{M}_{ii,\alpha\beta}^{tt}= \mu^t \delta_{\alpha\beta} ,\quad \textrm{M}_{ii,\alpha\beta}^{rr}= \mu^r \delta_{\alpha\beta}, \nonumber \\
&\textrm{M}_{ii,\alpha\beta}^{tr}= 0,\qquad\quad \textrm{M}_{ii,\alpha\beta}^{rt}= 0,
\end{align}
where $i=1\dots N$, $\delta_{\alpha\beta}$ denotes the Kronecker delta, and $\alpha,\beta=x,y$ label the different Cartesian coordinates.
Here, we introduced the translational and rotational mobility coefficients
\begin{equation}
 \mu^t=\frac{1}{6\pi\eta a}, \quad \mu^r=\frac{1}{8\pi\eta a^3},
\end{equation}
with $a$ denoting the hydrodynamic radius of the swimmer body and $\eta$ the viscosity of the surrounding fluid.
The remaining components of $\mathds{M}$ are given by \cite{dhont1996introduction}
\begin{align}
\textrm{M}_{ij,\alpha\beta}^{tt}(\vr)&= \mu^t \Bigl[ \frac{3a}{4r}\left( \delta_{\alpha\beta}+ \frac{r_\alpha r_\beta}{r^2} \right) \nonumber \\
&\qquad+\frac12{\left(\frac{a}{r}\right)}^3 \left(\delta_{\alpha\beta}-3 \frac{r_\alpha r_\beta}{r^2}\right) \Bigr], \\
\textrm{M}_{ij,\alpha\beta}^{rt}(\vr)&= \textrm{M}_{ij,\alpha\beta}^{tr}= \mu^r {\left(\frac{a}{r}\right)}^3 \sum_\gamma \epsilon_{\alpha\gamma\beta}\ r_\gamma, \\
\textrm{M}_{ij,\alpha\beta}^{rr}(\vr)&={}-\mu^r \frac12{\left(\frac{a}{r}\right)}^3 \left(\delta_{\alpha\beta}-3 \frac{r_\alpha r_\beta}{r^2}\right)
\end{align}
for $i\neq j$ ($i,j=1\dots N$), $\vr= \vR_j-\vR_i$, $r=|\vr|$, and $\epsilon_{\alpha\gamma\beta}$ the Levi-Civita tensor.

\begin{figure}[]
\centering
  \includegraphics[width=8.6cm]{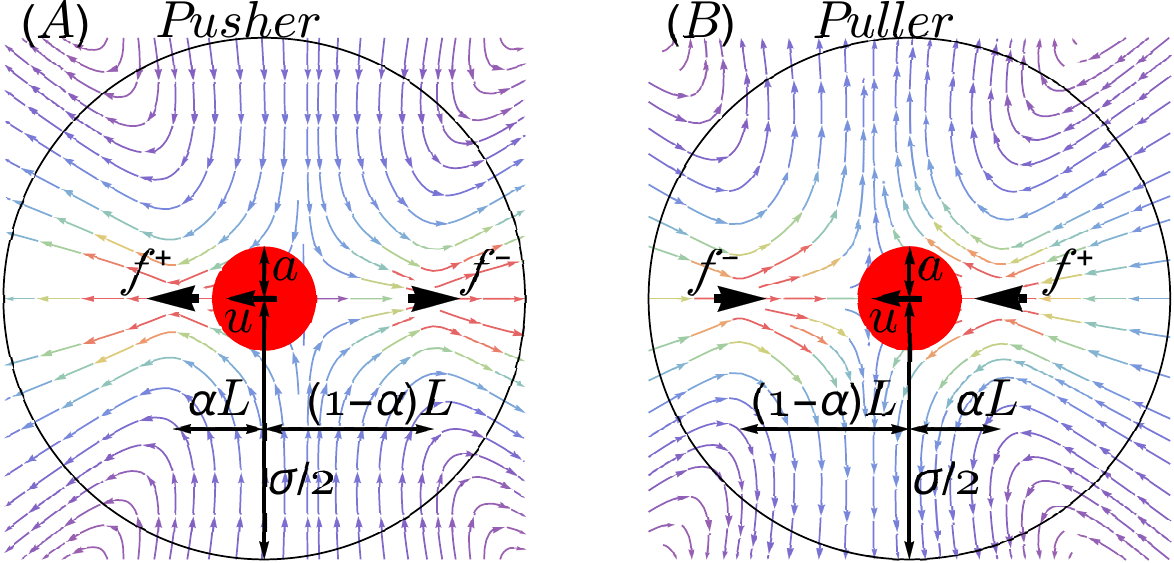}
  \caption{Geometry of our (A) pusher ($f>0$) and (B) puller ($f<0$) model microswimmers with the direction of net motion denoted by $\vu$.
  The spherical swimmer body of hydrodynamic radius $a$ is convected by the flow field indicated by field lines and arrows and induced by the two forces positioned at  $\pm\alpha L \vu$ and $\mp (1-\alpha) L \vu$ from the center of the sphere.
  The radius of the effective steric repulsion, see Eq.~(\ref{ster_pot}), is indicated by $\sigma/2$.
  Color of background arrows from purple (low) to red (high) indicates the local intensity of the flow field.}
  \label{figure01}
\end{figure}

So far, we have described passive particles interacting hydrodynamically with each other.
Now we include self-propulsion.
For this purpose, two point-like force centers are rigidly connected to each swimmer body, see Fig.~\ref{figure01}.
The two force centers are separated by a distance $L$ and exert on the fluid two oppositely oriented forces of equal magnitude along the symmetry axis of each swimmer.
Since the force centers are located at different distances from the sphere, the resulting flow field leads to a transport of the swimmer body in the self-induced fluid flow \cite{baskaran2009statistical, menzel2016dynamical, hoell2017dynamical}.
In principle, the rigid swimmer bodies affect the self-induced flow field \cite{adhyapak2017flow}.
Here, we do not include this effect.
That is, we only address the situation  to lowest order in the length scale $a/L$ \cite{hoell2017dynamical}.

The two active forces of the $i$-th swimmer are parametrized as
\begin{equation}
 \bm{f}_i^+= |f|\vu_i,\quad \bm{f}_i^-=-|f|\vu_i,
\label{active_f}
\end{equation}
with their centers located at the positions
\begin{equation}
 \vR_i^+=\vR_i +\frac{f}{|f|}\alpha L \vu_i,\quad \vR_i^-=\vR_i-\frac{f}{|f|}(1-\alpha) L \vu_i,
 \label{active_sites}
\end{equation}
respectively.
Here, $\alpha \in \ ]a/L,\ 1-a/L[$ quantifies the asymmetry in the propulsion mechanism.
The case of $\alpha=0.5$ recovers the symmetric ``shaker'' configuration \cite{baskaran2009statistical}.
Moreover, following Eq.~(\ref{active_sites}), the sign of $f$ determines whether the swimmer is a pusher or a puller, i.e., whether it pushes the fluid outward or pulls the fluid inward along the symmetry axis.
To calculate the effect of the active forces of swimmer $j$ on the motion of swimmer $i$, we use the  mobility matrices of components \cite{menzel2016dynamical, hoell2017dynamical}
\begin{align}
{\mu}_{\alpha\beta}^{tt}(\vr)&= \frac{1}{8\pi\eta r}\left( \delta_{\alpha\beta} +\frac{r_\alpha r_\beta}{r^2}\right) \nonumber \\
&\ \ \ + \frac{a^2}{24\pi\eta r^3}\left( \delta_{\alpha\beta} - 3 \frac{r_\alpha r_\beta}{r^2} \right), \\
{\mu}_{\alpha\beta}^{rt}(\vr)&= \frac{1}{8\pi\eta r^3} \sum_\gamma \epsilon_{\alpha \gamma \beta}\ r_\gamma .
\end{align}
Using these expressions, we obtain the components of the active mobility matrix $\mathds{A}$ in Eq.~(\ref{v_determ}) as \cite{menzel2016dynamical, hoell2017dynamical}
\begin{align}
&{\textrm{A}}_{ij,\alpha\beta}^{tt}= f\left[\mu^{tt}_{\alpha\beta}(\vr_{ij}^{+}) -\mu^{tt}_{\alpha\beta}(\vr_{ij}^{-})\right], \\
&{\textrm{A}}_{ij,\alpha\beta}^{rt}=  f\left[\mu^{rt}_{\alpha\beta}(\vr_{ij}^{+}) -\mu^{rt}_{\alpha\beta}(\vr_{ij}^{-})\right], \\
&{\textrm{A}}_{ij,\alpha\beta}^{tr}= {\textrm{A}}_{ij,\alpha\beta}^{rr} = 0.
\end{align}
$\vr_{ij}^{\pm}=\vR_j^\pm-\vR_i$ is the vector connecting the $j$-th $\pm$ active force site to the center of particle $i$.
The elements of $\mathds{A}^{rt}$ and $\mathds{A}^{rr}$ vanish because the propulsion forces are aligned with and are located on the symmetry axis of the swimmer and, thus, exert no active torque \cite{hoell2017dynamical}.

In the case of extremely diluted (i.e., non-interacting) swimmers, their self-propulsion speed $v_{0}$ follows as
\begin{equation}
 v_{0} = \frac{f \mu^t a}{2L}\left[ 3\frac{1-2\alpha}{\alpha(1-\alpha)} -\frac{a^2}{\alpha^3 L^2} +\frac{a^2}{(1-\alpha)^3 L^2}  \right].
\end{equation}
To position the force centers outside of the swimmer body, we require $\alpha \in \ ]a/L,\ 1-a/L[$.

Finally, our swimmers sterically interact with each other via the pair potential of the generalized exponential model of index 4 (GEM-4) \cite{archer2014solidification}
\begin{equation}\label{ster_pot}
 V^{st}(\vr) = \epsilon_{0} \exp\left(-\frac{{|\vr|}^4}{{\sigma}^4} \right),
\end{equation}
where $\epsilon_{0}$ and $\sigma$ measure, respectively, strength and range of the steric repulsion.
Although the steric interaction is soft, we indicate by $\sigma$ the size of the swimmers.
In the following, for convenience, we use $v_0$ as the unit of measure of velocities.
We set $a=\sigma/4\sqrt{3}$, $L=\sigma/2$, $\alpha=0.3$, and swimming forces $|f|\simeq 2.41 f_0$.
Distances, times, and forces are measured in multiples of $\sigma$, $t_0=\sigma/v_0$, and $f_0=v_0/\mu^t$, respectively.

Moreover, to compare our study with other theoretical investigations as well as with experimental results, we introduce the following dimensionless numbers.
First, the P\'eclet number
\begin{equation}\label{peclet}
 \Pe = \frac{v_0 \sigma}{\mu^t k_BT}
\end{equation}
quantifies the strength of self-propulsion with respect to Brownian diffusion.
Furthermore the area fraction is given by
\begin{equation}
\phi = \frac{N\pi\sigma^2}{4A},
\label{phieq}
\end{equation}
and the fraction of overall pushers by $\mixr=N_A/N$.
Thus, we indicate by $\mixr=0$ and $\mixr=1$ pure monodisperse systems of pullers and pushers, respectively.
Finally, unless specified otherwise, all of the following results are obtained for simulations with a total of $N=1024$ particles.
This, together with Eq.~(\ref{phieq}) and for a given area fraction $\phi$, sets the area of our periodic square box ${A}$. 



\section{Results}\label{results}
\begin{figure}[]
\centering
  \includegraphics[width=8.6cm]{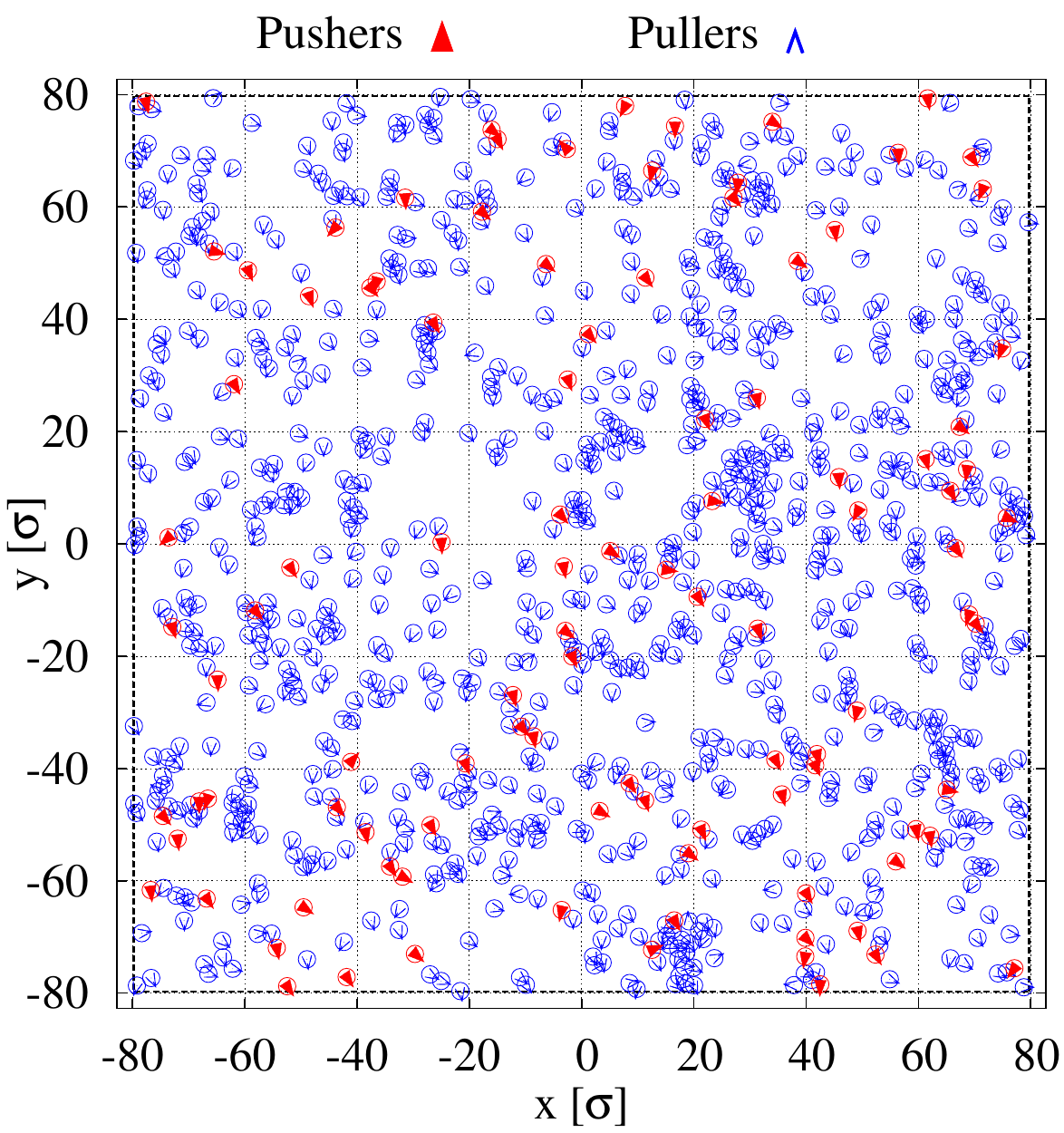}
  \caption{Example snapshot of one of our simulations of binary pusher-puller mixtures of active microswimmers, here at $\phi = 0.0316$, $\Pe=\infty$ ($k_BT=0$), and $\mixr=0.1$. The depicted state corresponds to a value of the polar order parameter of $P\simeq 0.78$.
  The orientations $\vu_i$ of each swimmer are indicated by the arrow hats. For better visibility, the sizes of the simmers have been enlarged.
  The total number of swimmers is $N=1024$, $N_A=103$ of which are pushers (larger filled red arrow hats) and $N_B=921$ of which are pullers (smaller empty blue arrow hats).
  Moreover, the black dashed square delimits the simulation box.}
  \label{figure_snapshot}
\end{figure}

\begin{figure}[]
\centering
  \includegraphics[width=8.6cm]{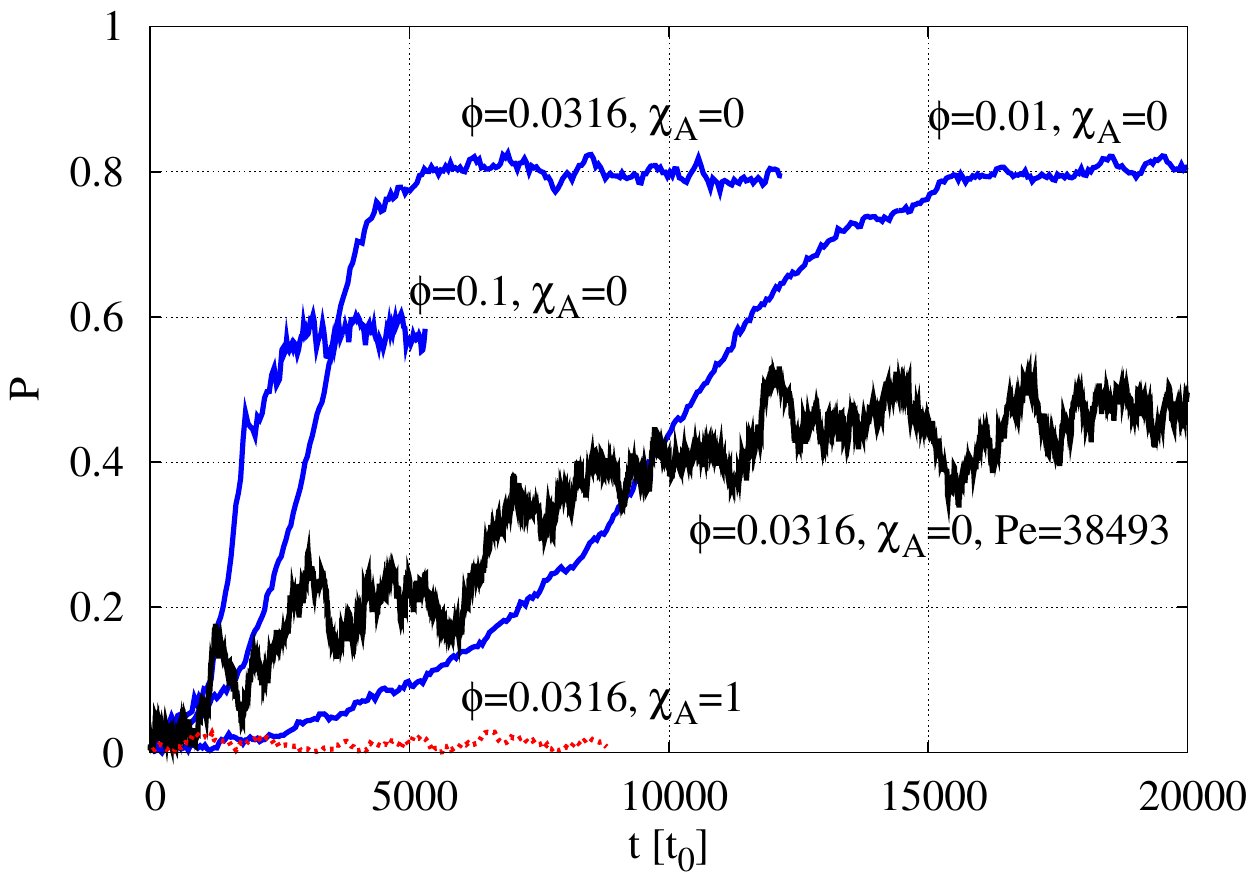}
  \caption{Time evolution of the polar order parameter $P(t)$ for different area fractions $\phi$, P\'eclet numbers $\Pe$, and relative amounts of pushers $\mixr$.
  The curves, unless specified otherwise by the respective labels, are obtained from simulations with $P(t=0)=0$, $\Pe=\infty$ ($k_BT=0$), and $\mixr=0$.
  The total number of active microswimmers is $N=1024$, except for the case of $\Pe=38493$ (in black) comprising only $N=225$ swimmers because of the higher computational cost.}
  \label{figure02}
\end{figure}

In our simulations, a suspension of active microswimmers can spontaneously develop collective motion into a common direction, see the example snapshot in Fig.~\ref{figure_snapshot}.
To describe the degree of such collective orientational ordering quantitatively, we define the global polar order parameter
\begin{equation}
P(t)=\left| \frac1N \sum_{i=1}^N \vu_i(t) \right|,
\end{equation}
which is equal to $1$ in the case of complete polar alignment of all swimmers and $0$ if the orientations do not show a net global polar order.
In all our simulations, we start from an initial configuration of isotropically distributed orientations, implying $P(t=0)=0$.
In agreement with the results obtained via a Lattice-Boltzmann scheme in Ref.~\cite{alarcon2013spontaneous}, and as shown in Fig.~\ref{figure02}, suspensions of only pullers spontaneously develop a steady polar order, which here seems to saturate around $P \sim 0.8$.
In the case of only pushers, instead, we in our system do not observe the polar order parameter to spontaneously increase; moreover, if initialized by an aligned state $P(t=0)=1$, $P(t)$ quickly decays to almost zero.

The overall area fraction $\phi$ affects the dynamics of developing ordered collective motion.
At low area fractions, e.g., $\phi=0.01$ in Fig.~\ref{figure02}, the swimmers eventually reach an equally high amount of alignment as for $\phi\sim 0.03$, but reaching this value takes a noticeably longer time.
The ordering process involves the induced flow fields acting on the other swimmers.
Lower area fractions imply larger interparticle distances, weaker hydrodynamic interactions, and longer time needed for the swimmers to develop the collective behavior.
At higher area fractions, instead, the time necessary to reach the steady state further decreases, see $\phi=0.1$ in Fig.~\ref{figure02}.
The attained orientational order, however, is lower, presumably, because for denser systems collisions between the swimmers become more relevant and affect the overall order

Mostly, the results that we report here were obtained at vanishing temperature $k_BT=0$, i.e., for infinite P\'eclet number $\Pe=\infty$.
In all considered cases, in which we examined the influence of finite temperature, we found it to lower the limiting value of $P(t)$ and increase its fluctuations, see Fig.~\ref{figure02}.

\begin{figure}[]
\centering
  \includegraphics[width=8.6cm]{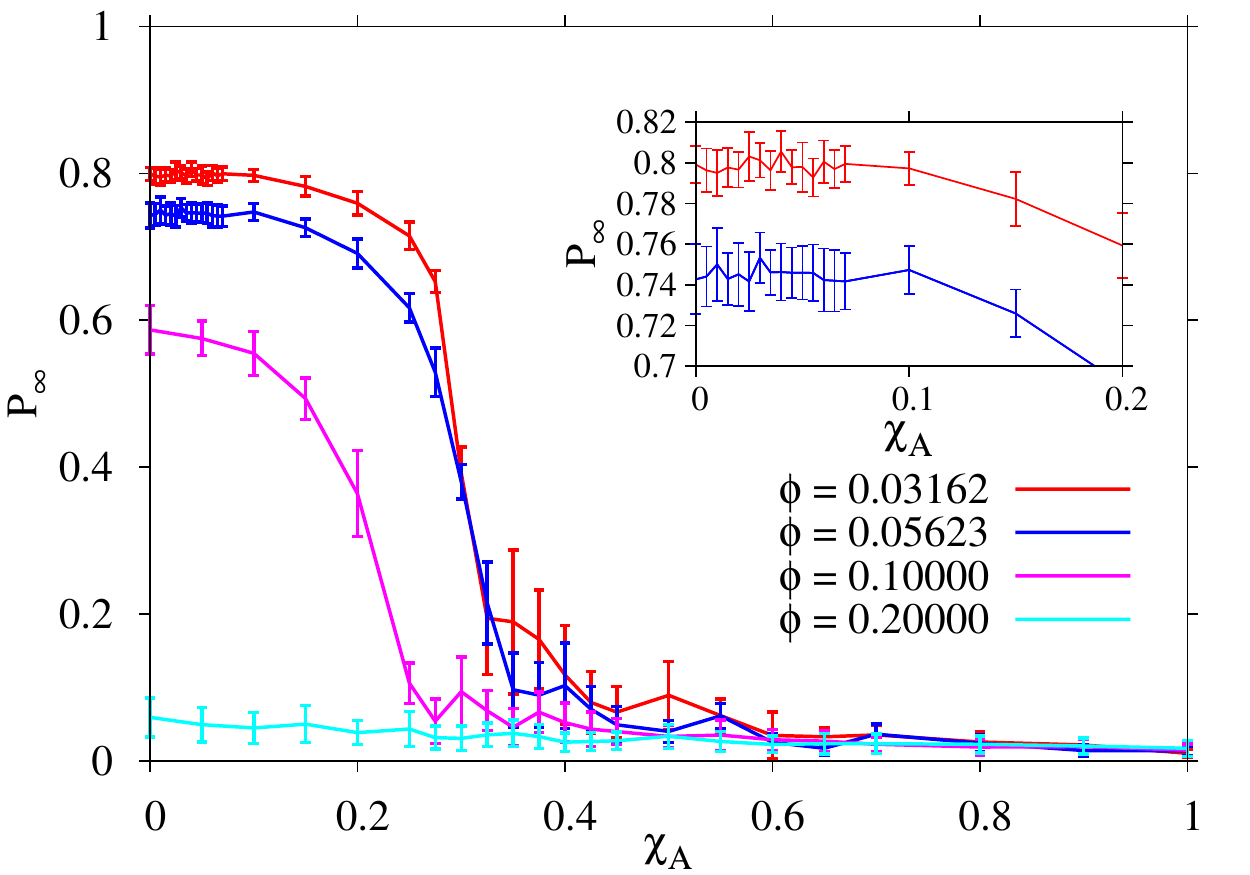}
  \caption{Polar order parameter $P_\infty$ in the stationary collective state for increasing pusher-puller mixing ratio $\mixr$ and different area fractions $\phi$ ($k_BT=0$, $\Pe=\infty$).
  Lines and bars represent averages and standard deviations, respectively, over sampling intervals in the stationary regimes as displayed in Fig.~\ref{figure02}.
  Inset: zoom of the initial behavior at low $\mixr$.}
  \label{figure03}
\end{figure}

We now move on to the central concern of our study, i.e., the collective behavior of pusher-puller mixtures.
For this purpose, we vary the fraction of pushers $\mixr = N_A/N$ from $0$ to $1$.
We sample the average polar order parameter in the stationary state, i.e.,
\begin{equation}
P_\infty= \frac1M \sum_{n=1}^{M} P(t_n)
\end{equation}
with $M > 3000$.
Sampling is performed over a time interval $t\in [t_1,t_M]$ in the long-time regime, for which a stationary state has been reached.
The effect of increasing mixing ratio $\mixr$ for different area fractions is shown in Fig.~\ref{figure03}.
As mentioned above, high area fractions (see $\phi=0.2$ in Fig.~\ref{figure03}) hinder the orientational ordering of the swimmers regardless of the swimmer species.
On the contrary, at low to intermediate area fractions, collective motion spontaneously emerges for $\mixr=0$ and is more or less preserved even upon introduction of relatively large amounts of pushers.
Even up to a total of $\sim 30\%$ of pushers, see the curves for $\phi= 0.03162$ and $\phi= 0.05623$ in Fig.~\ref{figure03}, $P_\infty$ remains as high as $0.4$, indicating still a significant degree of alignment.
As $\mixr$ further increases beyond this point, $P_\infty$ quickly decays to zero and the absence of polarly ordered collective motion in our pure pusher suspensions ($\mixr=1$) is recovered.

\begin{figure}[]
\centering
  \includegraphics[width=8.6cm]{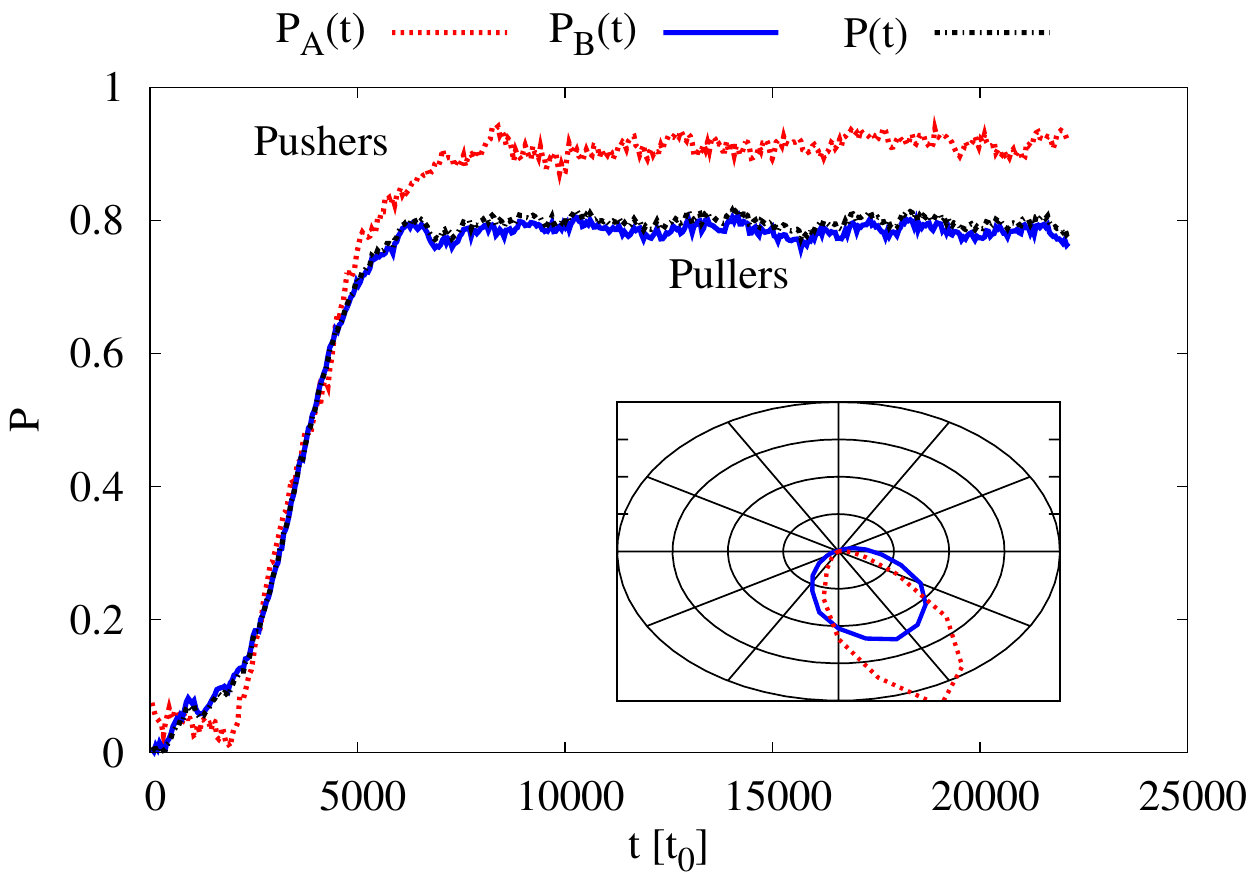}
  \caption{Time evolution of the polar order parameter $P(t)$ in a suspension of pullers with a 10\% doping by pushers ($\mixr=0.1$), area fraction $\phi=0.03162$, and $\Pe=\infty$ ($k_BT=0$).
  The dotted (red) line denotes the polar order solely of the pusher microswimmers, $P_A(t)$.
  Solid (blue) and dash-dotted (black) lines correspond to the polar order parameter of the pullers, $P_B(t)$, and of the whole collection, $P(t)$, respectively.
  Inset: polar distribution of the swimming orientations of pullers (solid, blue line) and pushers (dotted, red line) in the stationary regime.}
  \label{figure04}
\end{figure}

Remarkably, when a large set of pullers is doped with a small amount of pushers, the latter are observed to align themselves along the collective direction of motion more tightly than the surrounding pullers.
To illustrate this behavior, we show in Fig.~\ref{figure04} the polar order parameters for the two species separately, $P_A(t)$ for pushers and $P_B(t)$ for pullers.
In the stationary regime, we find $P_A(t)>P_B(t)$.
As a consequence of this higher degree of alignment, the distributions of the pusher and puller swimming orientations (see the inset of Fig.~\ref{figure04}) are centered on the same direction, but the pusher distribution is narrower.
Even for $\mixr$ as high as $0.3$, we found the polar order parameter $P_A(t)$ to be systematically higher than $P_B(t)$.

We remark that an increased orientational ordering and mutual support in collective motion by interactions between different species in a binary mixture of self-propelled particles has been previously reported in a ``dry'' system \cite{menzel2012collective}, analyzing a variant of the Vicsek model \cite{vicsek1995novel}.
In our case, such an effect of mutual support in orientational ordering would need to result from the presence of the hydrodynamic interactions due to the self-induced flow fields.
In the inset of Fig.~\ref{figure03}, we enlarge the curves for elevated polar order at low fractions of pushers $\mixr$.
Whether also the overall polar orientational order increases by the initial addition of pushers at low values of $\mixr$ cannot be statistically resolved by our present means.
This question needs further clarification in the future.

\begin{figure}[]
\centering
  \includegraphics[width=8.6cm]{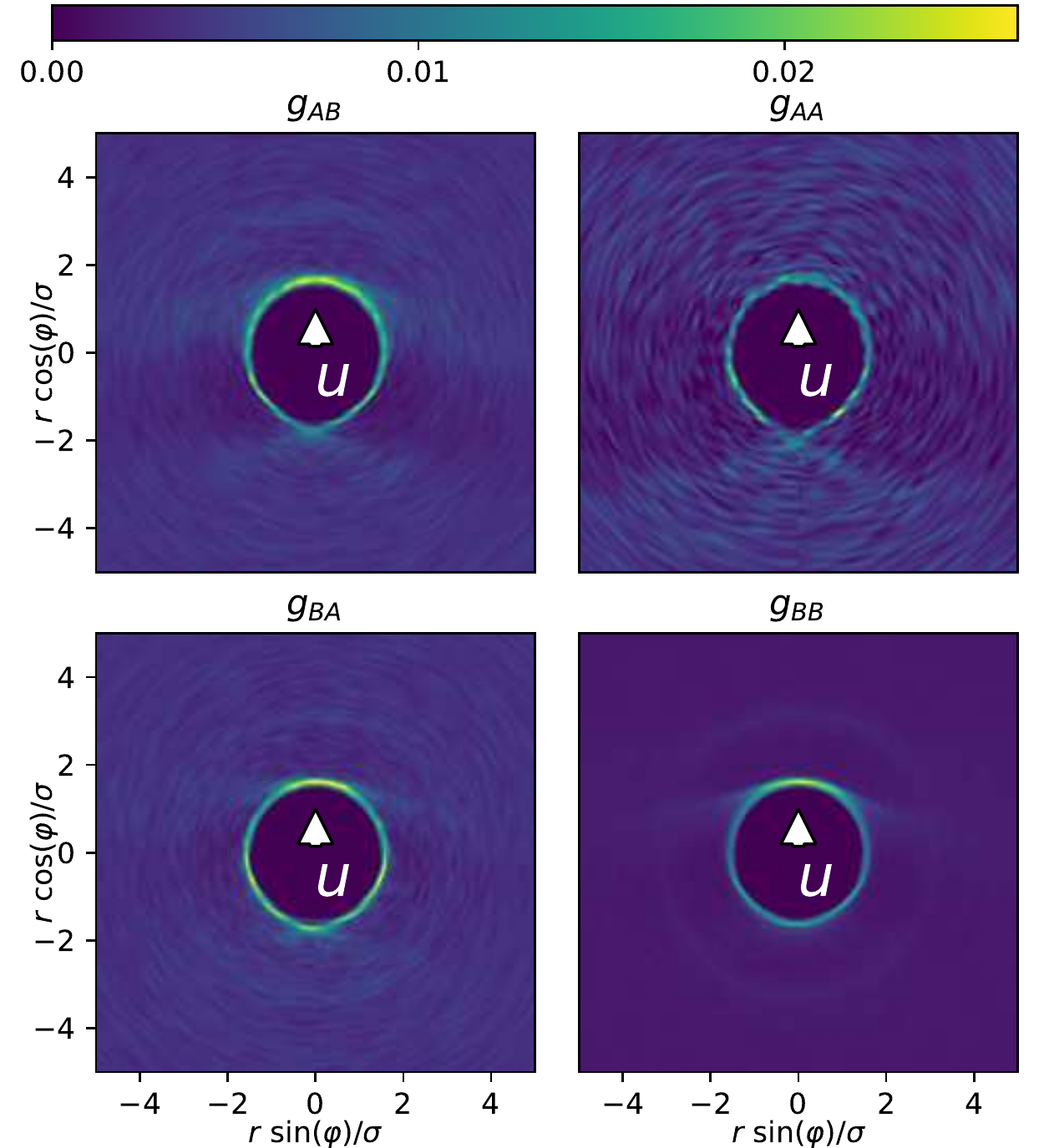}
  \caption{Pair distribution functions $g_{{XY}}(r,\varphi)$ (X,Y=A,B with A for pushers and B for pullers) related to the probability to find a swimmer of species Y in the $\varphi$-direction and at distance $r$ from a swimmer of species X centered at the origin and pointing to the top.
  The data are sampled in the stationary state of a simulation at $\phi=0.03162$, $k_BT=0$, and $\mixr=0.1$.
  $g_{{AB}}$, $g_{{BA}}$, and $g_{{AA}}$ have been rescaled for better visibility and their maximum intensity is $0.0125$.
  The overall $g(r,\varphi)$ for the whole system is basically indistinguishable from $g_{{BB}}(r,\varphi)$.}
  \label{figure05}
\end{figure}

Partial answer to this question can be obtained by evaluating the different pair distribution functions \cite{haertel2018three} $g_{{XY}} (r,\varphi)$, with X and Y either A (pushers) or B (pullers).
$g_{{XY}} (r,\varphi)$ represents the probability to find a swimmer of species Y around a swimmer of species X at distance $r$ and in direction $\varphi$ with respect to the swimmer orientation of X, see Fig.~\ref{figure05}.
The amount of doping by pushers in Fig.~\ref{figure05} was $\mixr=0.1$ and the overall total pair distribution function is virtually identical to $g_{{BB}}(r,\varphi)$.

The functions $g_{XY}(r,\varphi)$ feature a ring around the center, most likely due to the soft steric interaction introduced in Eq.~(\ref{ster_pot}), which was cut at $r\sim 2\sigma$.
$g_{BB}(r,\varphi)$ shows a central maximum at the front, which is presumably related to collisions between the self-propelled swimmers. 
Interestingly, $g_{{AB}}(r,\varphi)$ features three distinct maxima: one at the front and two lateral ones at $\varphi \approx \pm 3\pi/4$.
This may indicate a preferred arrangement for pushers when surrounded by a majority of pullers, namely, one or more pullers in front of each pusher and two behind at $\varphi \approx \pm 3\pi/4$. 
A similar triangular-like configuration is found for pullers when considering the probability to find a nearby pusher, see the function $g_{{BA}}(r,\varphi)$.
Remarkably, there is no pronounced maximum at the front for $g_{{AA}}(r,\varphi)$ for the pusher-pusher spatial correlation.
This may reflect the propulsion mechanism associated with the ejection of fluid along this axis.
Such flows will counteract the mutual approach of two pushers along this axis.



\section{Conclusions}\label{conclusions}
To summarize, we employed particle-resolved simulations to address the behavior of binary mixtures of self-propelled particles of different propulsion mechanisms (pushers and pullers).
The effect of mutual support between the two species concerning the polar orientational ordering of their propulsion directions was analyzed.
So far, this question of mutual inter-species coupling has been investigated within a variant of the famous Vicsek model for ``dry'' self-propelled particles \cite{menzel2012collective}.
Here, we have explicitly included the contribution of hydrodynamic interactions to the collective orientational behavior.

Via our minimal hydrodynamic microswimmer model, we can readily realize both pusher- and puller-like propulsion mechanisms.
In agreement with previous studies \cite{alarcon2013spontaneous}, we observe the spontaneous polar orientational ordering of pure monodisperse suspensions of pullers, while no polar ordering could be found in our monodisperse suspensions of pushers.
Furthermore, we point out that increased area fraction or temperature counteract the polarly ordered collective motion.
We remark that at very low area fractions the swimmers weakly interact and a common orientation could not be reached within observable times.

By doping a system of pullers even with significant amounts of pushers (up to $30\%$) the overall polar collective motion is largely preserved.
Surprisingly, we find that the polar ordering of pushers in this case is higher than the overall polar orientational order in the rest of the system.
Such an effect is possibly connected to some preferred spatial arrangement of the pushers relatively to the surrounding pullers.
One hint to support the existence of such preferred arrangements can be inferred from the inter-species pair distribution functions.
Further work is necessary in the future to determine the mechanism that drives the pushers into a more ordered state than the enclosing pullers.
A way to shed further light onto the internal structure of such mixtures could result from more analytical investigations, based, for instance, on dynamical density functional theory \cite{menzel2016dynamical, hoell2017dynamical}.

Several developments may follow on the basis of the present study.
First, polydispersity in size of both pushers and pullers could be considered, as well as different continuous combinations of the parameters $\alpha$ and $f$ related to propulsion efficiency and activity.
Moreover, an additional doping by passive particles should be addressed.
The effect of using other microswimmer models \cite{babel2016dynamics,najafi2004simple} could likewise be assessed in subsequent investigations.
In this way, a large set of parameters is to be explored to devise mixtures of different active and passive particles to adjust at will the structural and dynamic properties of the system.
Achieving tunable degrees of alignment for specific subsets of active particles could, for instance, allow to modify the transport properties or selectively separate the different species.
Via improved particle-resolved simulations, we hope to gain a better understanding and to develop elaborated predictions on the dynamic and structural behavior of real active systems.

\begin{acknowledgments}
\GP{This paper is dedicated to Daan Frenkel on the occasion of his 70th birthday.}
The authors thank M.~Puljiz and C.~Hoell for helpful discussions and the Deutsche Forschungsgemeinschaft for support of this work through the SPP 1726, grant nos.~LO 418/17 and ME 3571/2.
\end{acknowledgments}

\normalem


\end{document}